\documentclass[aps,prl,reprint]{revtex4-1}
\usepackage[pdftex]{graphicx}
\usepackage{times}
\usepackage{bm}
\usepackage{multirow}

\begin{document}

\title{On the problem of proton radius puzzle}
\author{D. Solovyev}
\email{solovyev.d@gmail.com}

\affiliation{Department of Physics, St.Petersburg State University, St.Petersburg, 198504, Russia}

\begin{abstract}
This paper concerns the most intriguing question of modern atomic physics: determination of the proton root-mean-square (rms) charge radius ($r_p$). This problem was announced by the difference in $r_p$ values extracted from the measurements of transition frequencies in hydrogen ($H$) atom and the Lamb shift in muonic hydrogen ($\mu H$) atom. In particular, it was found that the value of proton charge radius extracted from the $\mu H$ experiment is about $4\%$ smaller than the value given by the hydrogenic experiments. For the decade a lot of theoretical efforts were devoted to the search of 'new physics' on the basis of such deviation. In this paper the analysis of proton charge radius determination in the hydrogen atom is given. It is shown that the $r_p$ value extracted from the hydrogenic data can be found more close to the rms value given by the $\mu H$ experiment.
\end{abstract}
\maketitle

The question of the proton charge radius is still the subject of current theoretical and experimental investigations \cite{Carlson}. The root-mean-square charge radius, $r_p$, has been determined by three experiments: first, by the electron-proton scattering \cite{Sick}, \cite{Blunden}; second, by the precision spectroscopy of atomic hydrogen \cite{Fischer} and, third, by pulsed laser spectroscopy measurements of the Lamb shift in muonic hydrogen \cite{Pohl}. The most accurate $r_p$ value with the uncertainty of 1 per cent is based mainly on atomic hydrogen experiments and calculations of bound-state quantum electrodynamics (QED) \cite{Eides}. The present value given by CODATA using only electronic spectroscopy data is $r_p=0.8758(77)$ fm \cite{Mohr-2016}. Therewith the value given by the electron-proton scattering is $0.879(8)$ fm. Thus, CODATA finds that the overall result $r_p=0.8775(51)$ fm. The problem called 'proton radius puzzle' has arised from the muonic hydrogen experiment: matching the theoretical calculations of the Lamb shift with the experimental data leads to $r_p=0.84184(67)$ fm \cite{Pohl, Anton}. This magnitude differs on 5.6 standard deviations from the CODATA value. This discrepancy constitutes one of the most attractive questions in connection with the search of 'new physics'; a lot of theoretical and experimental efforts were devoted to investigation of this problem. 

The main problem of the proton charge radius determination from the hydrogenic data consists in complexity of theoretical description of such experiments, whereas the measurements in muonic hydrogen are transparent and allow the direct comparison of experiment with theory. However, the $r_p$ values obtained from electron-proton scattering and spectroscopic measurements in hydrogen are close that gives a reason for the inclusion of the overall value in CODATA. Very recently the new value of proton charge radius was reported in \cite{H-exp}: $r_p = 0.8335(95)$ femtometer. This value was extracted from the measurement of $2s-4p$ transition frequency in hydrogen atom and diverges on $1\%$ approximately from the $\mu H$ data. This value was extracted from the measurement of $2s-4p$ transition frequency in hydrogen atom and diverges on $1\%$ approximately from the $\mu H$ data. Such satisfactory agreement was reached in the experiment accounting the quantum interference effect and hyperfine splitting of levels. The attempts to describe theoretically the spectroscopic measurements in hydrogen were performed in a series of works \cite{LSPS,can-LSPS,PRA-LSPS,Jent-Mohr} on the base of \cite{LKG, LGL}, where the nonresonant corrections (called quantum interference in \cite{H-exp}) were introduced, see also reviews \cite{Andr, ZSLP-review}. In particular, theoretical description of the $1s-2s$ transition frequency measurement in hydrogen atom was given in \cite{PRL-LSSP,Can-LSSP,PRA-LSSC}, where the nonresonant correction to the $1s-2p$ transition frequency in hydrogen atom was estimated with the account for the hyperfine splitting also. According to the results of \cite{H-exp} the nonresonant effects should be taken into account in spectroscopic measurements and, therefore, the theoretical re-analysis of the proton charge radius determination from the hydrogenic data is required.

Precision spectroscopy of $H$ atom allows the accurate determination of fundamental physical constants: the Rydberg constant $R_\infty$ and the proton charge radius. It is achieved by the the highly accurate theoretical calculation of energy levels and experiments reaching the 15 digits in accuracy for the $1s-2s$ transition frequency \cite{Parthey,Mat}. Theoretical evaluation of hydrogen energies is performed according to the formula
\begin{eqnarray}
\label{1}
E_{nlj} = R_\infty\left(-\frac{1}{n^2}+f_{nlj}\left(\alpha,\frac{m_e}{m_p},r_p\dots\right)\right),
\end{eqnarray}
where $n$, $l$ and $j$ are the principal, orbital and total angular momentum quantum numbers, respectively. $R_\infty=m_e \alpha^2 c/2h$ is the Rydberg constant ($c$ is the speed of light and $h$ is the Planck's constant), $m_e$ and $m_p$ represent the electron and proton masses. The function $f_{nlj}$ denotes all the possible corrections arising within the relativistic QED theory, see \cite{Mohr-2016}.

To determine the Rydberg constant and proton radius the theoretical results should be compared with the corresponding experimental data: $E_{nlj}-E_{n'l'j'}=\Delta E^{\rm exp}_{nlj-n'l'j'}$, where transition frequencies $\Delta E^{\rm exp}_{nlj-n'l'j'}$ can be found in \cite{Mohr-2016}. To extract the Rydberg constant and proton charge radius the two independent transitions should be used. For this purpose the code reproducing the transition frequencies with an accuracy of 12 digits was written in {\it Wolfram Mathematica} software. In our calculations the values of fine structure constant $\alpha = 1/137.035999139$ and speed of light $с = 299792458$ $m/s$ were employed. Solving this system three times for the experimental value and plus/minus uncertainty, the root-mean-squared value and deviations can be defined via the expressions $x^{\rm rms}=\sqrt{\sum\limits_{i=1}^N\frac{x_i^2}{N}}$ and $\delta x^{\rm rms}=\sqrt{\sum\limits_{i=1}^N\frac{\delta x_i^2}{N-1}}$, respectively. The obtained data are listed in Table~\ref{tab:1}.
\begin{widetext}
\begin{center}
\begin{table}
\caption{The Rydberg constant, $R_\infty$, proton radius, $r_p$, and their deviations for the hydrogen atom. The second and third sublines represent the obtained values for the plus and minus experimental uncertainty, respectively. The rms values of Rydberg constant and proton radius with their deviations are given in the fourth subline. In the first column the pair of used transitions are listed.}\label{tab:1}
\begin{tabular}{|| l | c | c | c | c ||}
\hline
 Transitions & $R_\infty$ in $m^{-1}$ & $r_p$ in fm & $\delta R_\infty$ in $m^{-1}$ & $\delta r_p$ in fm \\ 
 
 \hline
(U. Sussex laboratory)  
\multirow{5}{*}{$2s_{1/2}-2p_{1/2}$, $1s_{1/2}-2s_{1/2}$} & $10973731.5690796$ & $0.927367$ &$0.0$ & $0.0$ \\

& $10973731.5697026$ & $0.980986$ &$0.0006229$& $0.053619$ \\

 & $10973731.5684566$ & $0.870452$ &$-0.0006230$& $-0.056915$\\

{\bf rms values}  & ${\bf 10973731.5691098}$ & ${\bf 0.927367}$ &$0.0006229$& $0.055292$\\

{\bf via diffrential}  & ${\bf 10973731.5690796}$ & ${\bf 0.927367}$ &$0.0006229$& $0.055167$\\

\hline
(Harvard laboratory)  
\multirow{5}{*}{$2s_{1/2}-2p_{1/2}$, $1s_{1/2}-2s_{1/2}$} & $10973731.5685501$ & $0.879224$ &$0.0$ & $0.0$ \\

& $10973731.5688304$ & $0.905031$ &$0.0002803$& $0.025807$ \\

& $10973731.5682697$ & $0.852637$ &$-0.0002804$& $-0.026587$\\

{\bf rms values}  & ${\bf 10973731.5685501}$ & ${\bf 0.879224}$ &$0.0002804$& $0.026199$\\

{\bf via diffrential}  & ${\bf 10973731.5685501}$ & ${\bf 0.879224}$ &$0.0002803$& $0.026185$\\

\hline
\multirow{5}{*}{$2s_{1/2}-2p_{3/2}$, $1s_{1/2}-2s_{1/2}$} & $10973731.568442$ & $0.869069$ &$0.0$ & $0.0$ \\

& $10973731.568816$ & $0.903702$ &$0.000374$& $0.034633$ \\

& $10973731.568068$ & $0.832998$ &$-0.000374$& $-0.036071$\\

{\bf rms values}  & ${\bf 10973731.568442}$ & ${\bf 0.869069}$ &$0.000374$& $0.035359$\\

{\bf via diffrential}  & ${\bf 10973731.568442}$ & ${\bf 0.869069}$ &$0.000374$& $0.035322$\\

\hline
\multirow{5}{*}{$2s_{1/2}-8s_{1/2}$, $1s_{1/2}-2s_{1/2}$} & $10973731.5684175$ & $0.866758$ &$0.0$ & $0.0$ \\

& $10973731.5686402$ & $0.887596$ &$0.0002227$& $0.020838$ \\

  & $10973731.5681949$ & $0.845407$ &$-0.0002226$& $-0.021351$\\

 {\bf rms values}  & ${\bf 10973731.5684175}$ & ${\bf 0.866758}$ &$0.0002227$& $0.021096$\\

 {\bf via diffrential}  & ${\bf 10973731.5684175}$ & ${\bf 0.866758}$ &$0.0002226$& $0.022408$\\

\hline
\multirow{5}{*}{$2s_{1/2}-8s_{1/2}$, $1s_{1/2}-3s_{1/2}$} & $10973731.568387$ & $0.860376$ &$0.0$ & $0.0$ \\

& $10973731.568579$ & $0.874942$ &$0.000192$& $0.014566$ \\

  & $10973731.568196$ & $0.845559$ &$-0.000191$& $-0.014817$\\

 {\bf rms values}  & ${\bf 10973731.568387}$ & ${\bf 0.860376}$ &$0.000192$& $0.014692$\\

 {\bf via diffrential}  & ${\bf 10973731.568387}$ & ${\bf 0.860376}$ &$0.000192$& $0.014689$\\

%
%
%
%
 
%
%
%
%

\hline
\multirow{5}{*}{$2s_{1/2}-4s_{1/2} - \frac{1}{4}(1s_{1/2}-2s_{1/2})$} & $10973731.568747$ & $0.897458$ &$0.0$ & $0.0$ \\

 $1s_{1/2}-2s_{1/2}$   & $10973731.569103$ & $0.929463$ &$0.000356$& $0.032005$ \\

 & $10973731.568391$ & $0.864268$ &$-0.000356$& $-0.03319$\\

{\bf rms values}  & ${\bf 10973731.568747}$ & ${\bf 0.897458}$ &$0.000356$& $0.032603$\\

 {\bf via diffrential}  & ${\bf 10973731.568747}$ & ${\bf 0.897458}$ &$0.000356$& $0.032575$\\

\hline
\multirow{5}{*}{$2s_{1/2}-4p_{3/2} - \frac{1}{4}(1s_{1/2}-2s_{1/2})$} & $10973731.568062$ & $0.832404$ &$0.0$ & $0.0$ \\

 $1s_{1/2}-2s_{1/2}$  & $10973731.568477$ & $0.872417$ &$0.000415$& $0.040013$ \\

  & $10973731.567647$ & $0.790368$ &$-0.000415$& $-0.042036$\\

{\bf rms values}  & ${\bf 10973731.568062}$ & ${\bf 0.832404}$ &$0.000415$& $0.041037$\\

{\bf via diffrential}  & ${\bf 10973731.568062}$ & ${\bf 0.832404}$ & $0.000415$ & $0.040975$\\

\hline
\multirow{5}{*}{$2s_{1/2}-8d_{3/2}$, $1s_{1/2}-2s_{1/2}$} & $10973731.568548$ & $0.879045$ &$0.0$ & $0.0$ \\

& $10973731.568766$ & $0.899147$ &$0.000218$& $0.020102$ \\

  & $10973731.568331$ & $0.858471$ &$-0.000217$& $-0.020574$\\

{\bf rms values}  & ${\bf 10973731.568548}$ & ${\bf 0.879045}$ &$0.000218$& $0.020339$\\

{\bf via diffrential}  & ${\bf 10973731.568548}$ & ${\bf 0.879045}$ & $0.000218$ & $0.020348$\\

\hline
\multirow{5}{*}{$2s_{1/2}-12d_{3/2}$, $1s_{1/2}-2s_{1/2}$} & $10973731.568297$ & $0.855289$ &$0.0$ & $0.0$ \\

& $10973731.568528$ & $0.877172$ &$0.000231$& $0.021883$ \\

  & $10973731.568066$ & $0.832831$ &$-0.000231$& $-0.022458$\\

{\bf rms values}  & ${\bf 10973731.568297}$ & ${\bf 0.855289}$ &$0.000231$& $0.022172$\\

{\bf via diffrential}  & ${\bf 10973731.568297}$ & ${\bf 0.855289}$ & $0.000204$ & $0.019577$\\

\hline
\multirow{5}{*}{$2s-4p$, $1s-2s$} & $10973731.568075$ & $0.833701$ &$0.0$ & $0.0$ \\

& $10973731.568171$ & $0.843058$ &$0.000096$& $0.009357$ \\

  & $10973731.5679798$ & $0.824237$ &$-0.000095$& $-0.009464$\\

{\bf rms values}  & ${\bf 10973731.568075}$ & ${\bf 0.833701}$ & $0.000096$ & $0.009411$\\

{\bf absolute error}  & ${\bf 10973731.568075}$ & ${\bf 0.833701}$ & $0.000095$ & $0.009399$\\
	\hline
	\hline
\end{tabular}
\end{table}
\end{center}
\end{widetext}

In particular, from Table~\ref{tab:1} follows that in all the listed transitions the Rydberg constant coincides with value given in \cite{H-exp} in 10 digits. The deviation is formed mostly by the experimental inaccuracy. The most coincidence with the result of \cite{H-exp} arises for the pair $1s-2s$, $2s_{1/2}-4p_{3/2}-\frac{1}{4}(1s_{1/2}-2s_{1/2})$: $R_\infty=10973731.568062$ m$^{-1}$ and $r_p=0.8324$ fm. However, the uncertainty defined as the rms value of deviation exceeds the experimental one in several times. In Table~\ref{tab:1} the values defined as the absolute error are listed in each fifth sublines also. This uncertainties were determined with the use of equation: $\delta R_\infty \frac{2 c}{\alpha^2}f(r_p)+\delta r_p \frac{2 c R_\infty}{\alpha^2}f'(r_p)=\delta\omega^{\rm exp}$, where $f'(r_p)$ denotes the corresponding derivative over $r_p$ of the level energies Eq. (\ref{1}) and $\delta\omega^{\rm exp}$ represents the experimental uncertainty.

Thus, the rms values of the Rydberg constant and proton charge radius can be found for all the pairs, except the $1s-2s$ and $2s-4p$, as $10973731.568503(270)$ m$^{-1}$ and $0.8745(253)$ fm, respectively. With the choice of close lying results, that corresponds to the pairs $1s-2s$, $2s_{1/2}-2p_{1/2}$ (Harvard); $1s-2s$, $2s_{1/2}-2p_{3/2}$; $1s-2s$, $2s-8s$ and $1s-2s$, $2s_{1/2}-8d_{3/2}$, we find
\begin{eqnarray}
\label{2}
r_p=0.8735(75)\,{\rm fm},
\\
\nonumber
R_\infty=10973731.568489(79)\,{\rm m^{-1}}.
\end{eqnarray}
These values are in good agreement with the results recommended by CODATA $10973731.568508(65)$ ${\rm m^{-1}}$ and $0.8759(77)$ fm \cite{Mohr-2016}. However, taking the lowest values of $R_\infty$ and $r_p$ (defined via the experimental value of frequency minus experimental uncertainty) for these transitions, the result $R_\infty=10973731.568216(98)$ m$^{-1}$, $r_p=0.8474(95)$ fm can be found. This proton charge radius value is in a good agreement with the $\mu H$-experiment data \cite{Pohl, Anton}. Thus, it can be concluded that the discrepancy of proton charge radii extracted from the $H$ and $\mu H$ experiments can be explained by the uncertainty of measurements in hydrogen atom. Moreover, for the pairs $1s-2s$, $2s-12d_{3/2}$ (lowest values); $1s-2s$, $2s-4p_{3/2}-\frac{1}{4}(1s-2s)$ (rms values); $1s-3s$, $2s-8s$ (lowest values); $1s-2s$, $2s-8s$ (lowest values); $1s-2s$, $2s-2p_{3/2}$ (lowest values) and $1s-2s$, $2s-2p_{1/2}$ (lowest values) one can find
\begin{eqnarray}
\label{3}
r_p=0.8403(79)\,{\rm fm},
\\
\nonumber
R_\infty=10973731.568143(81)\,{\rm m^{-1}}.
\end{eqnarray}
Here the proton charge radius coincides with the value given by $\mu H$ experiment $0.84095(39)$ fm \cite{Anton}, and the Rydberg constant lies within the error of \cite{H-exp}.

The case of $2s-4p$, $1s-2s$ pair deserves the separate consideration. To determine $R_\infty$ and $r_p$, the data from \cite{Mohr-2016} and $616520931626.8(2.3)$ kHz for the $2s-4p$ transition frequency found in \cite{H-exp} were used. The rms values of Rydberg constant and proton charge radius are $10973731.568075$ ${\rm m^{-1}}$ and $0.8337$ fm, respectively, that coincides with \cite{H-exp}.

Such displacement of $R_\infty$ and $r_p$ values in respect to the recommended by CODATA ($10973731.568508(65)$ ${\rm m^{-1}}$ and $0.8759(77)$ fm) was explained in \cite{H-exp} by the quantum interference effect. In the early paper by F. Low \cite{Low} it was pointed out that the description of spectral line is valid only up to a certain limit of accuracy which is defined by the nonresonant (NR) corrections. Theory of the NR corrections was developed in \cite{LKG,LGL} for $H$-like ions and the corresponding evaluation was prolongated to the hydrogen atom in \cite{LSPS,can-LSPS,PRA-LSPS,Jent-Mohr}. The main conclusion made in these works is that the nonresonant corrections set a principal limit for the accuracy of the resonance frequency measurements. 

\begin{figure}[hbtp]
	\centering
	\includegraphics[scale=0.15]{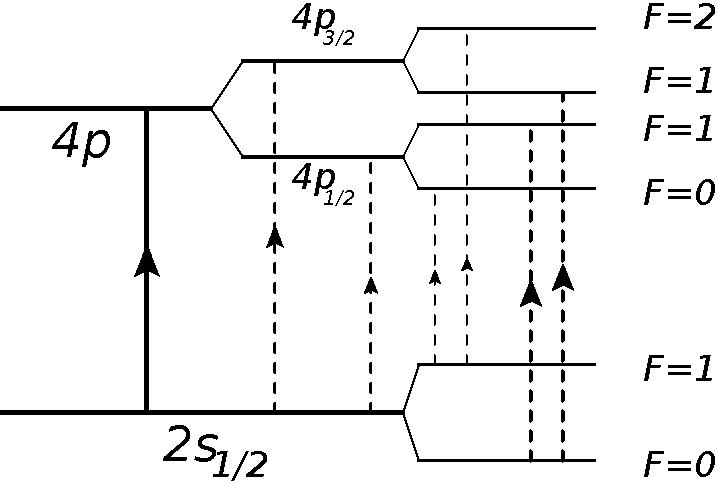}
	\caption{The fine and hyperfine structure of the $4p$ state in hydrogen atom are shown schematically. The corresponding transitions from hyperfine splitted $2s_{1/2}$ sublevels to the hyperfine splitted sublevels of the $4p_{1/2}$ and $4p_{3/2}$ atomic levels are illustrated.}
	\label{fig:1}
\end{figure}

The nonresonant correction to the differential cross-section with the account for the fine structure of levels was found in \cite{Jent-Mohr}. Nonetheless, this correction can be avoided by the measurement of 'gravity center' of spectral sublines. This procedure was applied in the experiment \cite{H-exp}, where authors have considered the fine and hyperfine structures of the $2s$ and $4p$ states in hydrogen atom, see Fig.~\ref{fig:1}.
Then the quantum interference occurs for the two transitions with equal quantum numbers: $2s_{1/2}^{F=0}\rightarrow 4p_{1/2}^{F=1}$ and $2s_{1/2}^{F=0}\rightarrow 4p_{3/2}^{F=1}$. Authors of \cite{H-exp} found that the $2s-4p$ centroid frequency should be shifted on $\delta\nu^{\rm exp}=-0.132552092 \cdot 10^9$ Hz and, as a consequence, the new values of the proton charge radius and Rydberg constant were determined. 

However, there is a non-resonant correction to the total cross-section arising due to the fine structure of levels which can not be avoided by this procedure \cite{PRA-LSPS}. The nonresonant correction is given by the expression $\delta_{\rm NR} = \frac{\Gamma^4_{a}}{16(\Delta E_f)^3}$, where $\Gamma_{a}$ represents the level width of state $a$ and $\Delta E_f$ is the energy of corresponding fine splitting. The rough estimation of this correction is of the order of the experimental accuracy of the $1s-2s$ transition frequency and, thus, is negligible. There is also the nonresonant correction arising with the account for the hyperfine splitting of levels, see \cite{PRL-LSSP}. The {\it quadratic} nonresonant correction of this type is
\begin{eqnarray}
\label{4}
\delta_{\rm NR} = \frac{\Gamma^4_{a}}{16(\Delta E_{hfs})^3}.
\end{eqnarray}
Then, with the use of experimental values $\Gamma_{4p}=20$ MHz and $\Delta E_{hfs}=-132,552.092$ kHz for the level widths and frequency splitting, the NR correction can be found as $\delta_{\rm NR} = -4293.78$ Hz. This correction has a opposite sign for the transitions $2s-4p_{1/2}$ and $2s-4p_{3/2}$. Thus, according to \cite{H-exp} the total asymmetry is $\delta\nu_{NR}=\frac{1}{3}\delta_{\rm NR}=1431.26$ Hz. Accounting for this shift leads to
\begin{eqnarray}
\label{5}
r_p=0.8395(93)\,{\rm fm},
\\
\nonumber
R_\infty=10973731.568135(96)\,{\rm m^{-1}}.
\end{eqnarray}

The same result can be achieved in other way: two transitions $2s-4p_{1/2}$ and $2s-4p_{3/2}$ in conjunction with $1s-2s$ transition can be analysed separately. Then with the account for the hyperfine splitting of levels one can write
\begin{eqnarray}
\label{6}
\Delta E^{\rm theor}\left(2s_{1/2}-4p_{1/2}\right) &=& \Delta E_{\rm exp}+\Delta_{hfs}^{2s^{F=0}_{1/2}}-\Delta_{hfs}^{4p^{F=1}_{1/2}},\qquad
\\
\nonumber
\Delta E^{\rm theor}\left(2s_{1/2}-4p_{3/2}\right) &=& \Delta E_{\rm exp}+\Delta_{hfs}^{2s^{F=0}_{1/2}}-\Delta_{hfs}^{4p^{F=1}_{3/2}},
\end{eqnarray}
where values of the hyperfine splitting can be taken from \cite{Horb}. Solving these systems of equations, the values $r_p^{(1/2)}=0.8272$ fm, $R^{(1/2)}_\infty=1097371.56800986$ m$^{-1}$ and $r_p^{(3/2)}=0.8369$ fm, $R^{(3/2)}_\infty=1097371.56810850$ m$^{-1}$ can be obtained, respectively. Then, according to \cite{H-exp}, we find $\frac{1}{3}r_p^{(1/2)}+\frac{2}{3}r_p^{(3/2)} = 0.8337$ fm and $\frac{1}{3}R^{(1/2)}_\infty+\frac{2}{3}R^{(3/2)}_\infty = 1097371.568076$ m$^{-1}$.

This procedure can be applied to other transitions. For example, for the pair $1s-2s$, $2s-4p_{1/2}-\frac{1}{4}(1s-2s)$ the solution is $r_p=0.9322(549)$ fm and $R_\infty = 10973731.569135(623)$ m$^{-1}$
. The lowest values are $r_p^{(1/2)}=0.8756$ fm and $R^{(1/2)}_\infty = 10973731.568512$ m$^{-1}$. Therefore, $r_p=\frac{1}{3}r_p^{(1/2)}({\rm lowest})+\frac{2}{3}r_p^{(3/2)} = 0.8468$ fm and $R_\infty = 10973731.568212$ m$^{-1}$. 
Employing Eq. (\ref{6}) for the Lamb shift (Harvard laboratory data), one can find $r_p=0.8490(271)$ fm and $R_\infty = 10973731.568232(280)$ m$^{-1}$. The corresponding magnitudes for the $1s-2s$, $2s-2p_{3/2}$ pair are $r_p^{(3/2)}=0.8691(353)$ fm, $R^{(3/2)}_\infty = 10973731.568442(374)$ m$^{-1}$. Combination $\frac{1}{3}r_p^{(1/2)}+\frac{2}{3}r_p^{(3/2)}$ yields $r_p=0.8624$ fm, $R_\infty = 10973731.568250$, and $r_p=\frac{1}{3}r_p^{(1/2)}+\frac{2}{3}r_p^{(3/2)}({\rm lowest}) = 0.8383$ fm and $R_\infty =10973731.568123$ m$^{-1}$. 

The combination of transitions with the different fine structure sublevels can be obtained with the use of weight coefficient $\frac{(2j+1)}{(2s+1)(2l+1)}$. Then for the frequencies $2s_{1/2}-np_{1/2}$ and $2s_{1/2}-np_{3/2}$ we obtain $2s-np = \frac{1}{3}(2s_{1/2}-np_{1/2})+\frac{2}{3}(2s_{1/2}-np_{3/2})$, whereas for the $2s_{1/2}-nd_{3/2}$ and $2s_{1/2}-nd_{5/2}$: $2s-nd = \frac{2}{5}(2s_{1/2}-nd_{3/2})+\frac{3}{5}(2s_{1/2}-nd_{5/2})$. The results with the account for the hyperfine structure for the transitions $2s-2p_{1/2}$, $2s-8d_{5/2}$ and $2s-12d_{5/2}$ are listed in first (upper) part of Table~\ref{tab:2} and in the second (lower) part of Table~\ref{tab:2} the combination of these transitions with the data from Table~\ref{tab:1} are given,the lowest values are listed in brackets.
\begin{widetext}
\begin{center}
\begin{table}
\caption{The hyperfine shift, Rydberg constant, $R_\infty$, proton radius, $r_p$, and their deviations with the account for the hyperfine shift. All the listed transitions were used in pair with the $1s-2s$ frequency. The second subline in each row represents the values obtained without the $\delta_{hfs}$.}\label{tab:2}
\begin{tabular}{|| l | c | c | c ||}
\hline
 Transitions & $\delta_{hfs}$, Hz &$R_\infty$, $m^{-1}$ & $r_p$, fm  \\ 
 
 \hline
\multirow{2}{*}{$2s_{1/2}-2p_{1/2}$} & $-|\Delta_{hfs}^{2s^{F=1}_{1/2}}|+|\Delta_{hfs}^{2p^{F=0}_{1/2}}|$, & ${\bf 10973731.568232(280)}$ & ${\bf 0.8490(271)}$  \\
 & $10208.6$ & $10973731.5685501(280)$ & $0.8792(262)$  \\
 

\hline
\multirow{2}{*}{$2s^{F=1}_{1/2}-8d_{5/2}$} & $-|\Delta_{hfs}^{8d^{F=2}_{5/2}}|+|\Delta_{hfs}^{8d^{F=3}_{5/2}}|$, & ${\bf 10973731.568057(168)}$ & ${\bf 0.8319(166)}$ \\
 & $-23766.6$ & $10973731.568681(168)$ & $0.8913(155)$ \\

\hline
\multirow{2}{*}{$2s^{F=1}_{1/2}-12d_{5/2}$} & $-|\Delta_{hfs}^{12d^{F=2}_{5/2}}|+|\Delta_{hfs}^{12d^{F=3}_{5/2}}|$, & ${\bf 10973731.568219(172)}$ & ${\bf 0.8478(167)}$ \\
 & $-7033.3$ & $10973731.568392(172)$ & $0.8643(163)$ \\

%
\hline
\hline
rms &  & $10973731.568169(80)$ & $0.8429(78)$ \\
	\hline
	\hline

$2s-2p$ & -- & $10973731.568250\,(10973731.568159)$ & $0.86238(0.8419)$ \\

  
\hline  
$2s-8d$  & -- & $10973731.568253\,(10973731.568167)$ & $0.85076(0.8425)$ \\
   
\hline
$2s-12d$  & -- & $10973731.568825\,(10973731.568158)$ & $0.85079(0.8418)$ \\

\hline
\hline
\end{tabular}
\end{table}
\end{center}
\end{widetext}

In particular, from Table~\ref{tab:2} follows that the rms values of the proton charge radius and the Rydberg constant are 
\begin{eqnarray}
\label{7}
r_p=0.8429(78)\,{\rm fm},
\\
\nonumber
R_\infty=10973731.568169(80)\,{\rm m^{-1}}.
\end{eqnarray}
The lowest value of the Rydberg constant and proton charge radius are $r_p=0.84207(33)$ fm, $R_\infty=10973731.568167(34)$ m$^{-1}$.

Concluding one can stay the point that the results for determination of the Rydberg constant and the proton charge radius from the hydrogenic experiments depend strongly on the experimental uncertainty \cite{Mohr-2016}. The uncertainty of frequency measurement leads to the rms deviation which exceeds the corresponding CODATA value in several times. Analysis of lowest values  gives the results more close to the $\mu H$ experiment and can be explained by the systematic error occuring for the measurements in hydrogen atom. At the same time, it is more believable that the experiment on muonic hydrogen atom is more dogmatic for the determination of the proton charge radius.

The analysis given in \cite{H-exp} for the $2s-4p$ transition reveals the necessity of the accounting for the hyperfine level structure and quantum interference effects that leads to the nonresonant corrections. Inclusion of the NR correction in the analysis of $2s-4p$ transition leads to results Eq. (\ref{5}). The relative difference between this value of the proton charge radius and $\mu H$ is about $0.2\%$. The results of calculations for the different transitions are listed in Table~\ref{tab:1} without the hyperfine shift of frequencies, and the results with use of procedure given in \cite{Anton} and \cite{H-exp} are presented in Table~\ref{tab:2}. In particular, from Tables~\ref{tab:1} and \ref{tab:2} follows that the $\mu H$ proton charge value can be restored from the hydrogenic data, see Eqs. (\ref{3}) and (\ref{7}).

\bibliography{mybibfile}

\end{document}